\RequirePackage{fix-cm}
\documentclass[onecolumn,epjc3]{svjour3}
\usepackage{mathtext,bbm,amsmath,amsfonts,amssymb,indentfirst,syntonly,graphicx}
\smartqed  
\RequirePackage{graphicx}
\RequirePackage{subfigure}
\RequirePackage{xcolor}

\def\bc{\begin{center}}
\def\ec{\end{center}}
\def\be{\begin{eqnarray}}
\def\ee{\end{eqnarray}}

\journalname{Eur. Phys. J. C}
\begin{document}

\title{Effects of spacetime anisotropy on the galaxy rotation curves}


\author{Zhe Chang \thanksref{addr1,addr2},
           Ming-Hua Li \thanksref{addr1},
           Xin Li \thanksref{addr1,addr2},
           Hai-Nan Lin \thanksref{e1,addr1},
           Sai Wang \thanksref{addr1}.
}
\thankstext{e1}{e-mail: linhn@ihep.ac.cn}
\institute{Institute of High Energy Physics, Chinese Academy of Sciences, 100049 Beijing, China\label{addr1}\and
           Theoretical Physics Center for Science Facilities, Chinese Academy of Sciences, 100049 Beijing, China\label{addr2}
}
\date{Received: date / Accepted: date}

\maketitle

\begin{abstract}
The observations on galaxy rotation curves show significant discrepancies from the Newtonian theory. This issue could be explained by the effect of the anisotropy of the spacetime. Conversely, the spacetime anisotropy could also be constrained by the galaxy rotation curves. Finsler geometry is a kind of intrinsically anisotropic geometry. In this paper, we study the effect of the spacetime anisotropy at galactic scales in the Finsler spacetime. It is found that the Finslerian model has close relations with the Milgrom's MOND. By performing the best-fit procedure to the galaxy rotation curves, we find that the anisotropic effects of the spacetime become significant when the Newtonian acceleration \(GM/r^2\) is smaller than the critical acceleration \(a_0\). Interestingly, the critical acceleration \(a_0\), although varies between different galaxies, is in the order of magnitude \(cH_0/2\pi\sim 10^{-10}~\rm{m\,\, s^{-2}}\).
\end{abstract}

\section{Introduction}\label{introduction}

The \(\rm \Lambda\)CDM model \cite{Weinberg2008} is well consistent with the present observations, such as the Wilkinson Microwave Anisotropy Probe (WMAP) \cite{WMAP7}, the Supernovae Cosmology Project \cite{Union2.1}, the Sloan Digital Sky Survey (SDSS) \cite{SDSS}, and so on. However, there are several evidences to challenge the \(\Lambda\)CDM model, see the reference \cite{Perivolaropoulos} for instance. For example, evidence has been claimed for the existence of a large-scale bulk flow by surveying the peculiar velocities of the clusters of galaxies \cite{Kashlinsky0809,Kashlinsky0910}. The bulk flow points to the direction \((l,b)=(283^\circ\pm14^\circ,12^\circ\pm14^\circ)\) with the velocity up to \(\sim1000~\rm{km/s}\) at the scale up to \(z\lesssim0.2\). By analyzing the quasar absorption spectra, Webb et al. found a large-scale variation of the fine structure constant \(\alpha\) \cite{WebbKMFCB2011}. The \(\alpha\) variation takes the spatial dipole form with a significant \(4.2\sigma\). Its direction is close to that of the bulk flow. In addition, the low multipoles of the CMB power spectrum align to a common direction approximately \cite{TegmarkOH2003,LandM2005}. The direction is also close to that of the bulk flow. Furthermore, Antoniou \& Perivolaropoulos showed a privileged direction \((l,b)=({309^\circ}^{+23^\circ}_{-3^\circ},{18^\circ}^{+11^\circ}_{-10^\circ})\) for the maximum acceleration of the universe by analyzing the dataset of type Ia supernovae (SNe Ia) \cite{AntoniouPerivolaropous2010} .

It is noteworthy that the directions mentioned above are close to each other at large scales, even though they origin from different astrophysical (or cosmological) observations. This result could be explained by hidden common systematic effects. However, it could also stem from one common physical issue. One possible physical explanation involves one large-scale privileged axis in the space, see reference \cite{AntoniouPerivolaropous2010}. For example, the anisotropic inflation model predicts a statistically anisotropic primordial power spectrum via a primordial vector field \cite{Lim,KannoSoda,Arianto,KoivistoMota,KohHu,Watanabe}. The vector field singles out a privileged direction which influences the evolution of early inflationary phase. Thus, the existence of the spatial privileged axis reveals the possible anisotropy of the spacetime.

Another challenge for the \(\Lambda\)CDM model involves the observations on Bullet Clusters \cite{Lee:2010hja}. In fact, the dark matter hypothesis originated from the observations on galaxy rotation curves. According to the Newtonian gravity, the rotational velocity is inversely proportional to the square root of distance from the center of a galaxy. However, the observed data often show an asymptotically flat rotation curve out to the furthest data points \cite{Rubin1980,Walter2008}. There are several ways to solve this problem, such as the dark matter \cite{Begeman1991,Persic1996,Chemin:2011mf}, the modified Newtonian dynamics (MOND) \cite{Milgrom1983a,Milgrom1983b}, the Horava\---Lifshitz (HL) theory \cite{Horava2009a,Horava2009b,Horava2009c}, and so on. The MOND and the HL theory could not also explain the mass discrepancy problem of the Bullet clusters. Later, Cardone et al. showed that the HL-theory is not in agreement with the observational data \cite{Cardone2010,Cardone2012}. Seemingly, the issue of galaxy rotation curves is irrelative to the anisotropy of the spacetime. The reason is that the 3-dimensional space--space rotations are often preserved for the galaxy rotation curves. Nevertheless, the time--space rotations (boosts) might be violated at the galactic scales. Thus, the galaxy rotation curves could provide a new platform to test the anisotropy of the spacetime. Or equivalently, the galaxy rotation curves could be explained by the spacetime anisotropy.

Finsler geometry gets rid of the quadratic constraint on the Finsler structure \cite{Book by Bao}. It admits certain privileged directions in the Finsler spacetime and permits less symmetries than the Riemann structure \cite{Finsler isometry by Wang,Finsler isometry LiCM}. Thus, Finsler geometry is a reasonable candidate to deal with the anisotropic properties of the spacetime. For example, the Finsler spacetime could potentially account for the Lorentz violation and the CPT violation \cite{A special-relativistic theory of the locally anisotropic space-time I,A special-relativistic theory of the locally anisotropic space-time II,A special-relativistic theory of the locally anisotropic space-time Appendix,recoiling D-branes02,DSR in Finsler,VSR in Finsler,Kostelecky_Finsler,ChangWangepjc 2012,ChangWangepjc 2013}. The cosmological acceleration could explained by the anisotropic Friedmann equation of motion in Finsler cosmology \cite{GVSR in Finsler cosmology}. The large-scale bulk flow could be explained by a Finslerian Zermelo navigation model \cite{Chang et al 2013a}. The variation of \(\alpha\) could be revealed by the anisotropic redshift in a Randers spacetime \cite{ChangLiWangepjc 2012}. Similarly, the Randers spacetime could also account for the privileged axis of the maximum acceleration in the Hubble diagram \cite{Chang et al 2013b}. A spatially anisotropic Finslerian model could account for the mass discrepancy problem of the Bullet cluster 1E0657-558 \cite{Lietal2012}.

In this paper, we propose a galactic test of the spacetime anisotropy via the galaxy rotation curves (or an explanation for the galaxy rotation curves in terms of the anisotropic spacetime). We first resolve the gravitational vacuum field equation in the anisotropic Finsler spacetime. At the Newtonian limit, we study the geodesic equations for particles cycling a gravitational source. The result shows that the modified dynamics is similar to that in the Newtonian dynamics. However, the spatial distance acquires certain anisotropic modifications from the Finsler structure. Thereafter, we call such a class of anisotropic Finsler models as the Finslerian MOND for convenience. The reason is that such Finsler models have close relations with the Milgrom's MOND as will be discussed latter. As is mentioned above, the galaxy rotation curves could be a new platform to test the anisotropy of the spacetime. The spacetime anisotropy at the galactic scales may be different from that at the large scales. We will fit the galaxy rotation curves with the Finslerian MOND to constrain the spacetime anisotropy at such galactic scales. Finally, the scale \(a_0\) would be obtained for the occurrence of the possible spacetime anisotropy.

The rest of the paper is organized as follows: In section \ref{sec:FMOND}, we give a derivation of the Finslerian MOND based on Finsler gravity. By resolving the Finsler gravitational vacuum equation and the Finsler geodesic equations, we obtain an anisotropic modification of the spatial distance from the Finsler structure. In section \ref{sec:MMOND}, we make a comparison between the Finslerian MOND and the Milgrom's MOND. We find that the Finslerian MOND is a kind of MOND with the interpolation function $\mu(x)$ different from that of the Milgrom's MOND. We use the Finslerian MOND to fit the observed rotation curve data to constrain the possible occurrence scale for the spacetime anisotropy in section \ref{sec:mass-profile}. Finally, discussions and conclusions are given in section \ref{sec:discussion}.

\section{The Finslerian MOND}\label{sec:FMOND}

Finsler geometry is a generalization of Riemannian geometry \cite{Book by Bao}. It is based on the so called Finsler structure $F$, which is a non-negative real function and has the property $F(x, \lambda y) \equiv \lambda F(x, y)$ for all $\lambda > 0$. Here, $x$ represents the position and $y\equiv dx/d\tau$ represents the velocity, and $\tau$ is the proper time. The Finsler metric is given as
\begin{equation}
g_{\mu\nu}\equiv\frac{\partial}{\partial y^{\mu}}\frac{\partial}{\partial y^{\nu}}\left(\frac{1}{2}F^2\right).
\end{equation}
The arc length in the $n$-dimensional Finsler space is given as
\begin{equation}
\int_s^r F(x^1,\cdots,x^n;\frac{dx^1}{d\tau},\cdots,\frac{dx^n}{d\tau})d\tau.
\end{equation}

Note that the Finsler gravity is covariant although the Finsler spacetime admits the anisotropy. Li et al. \cite{Lietal2012} introduced the gravitational vacuum field equation in Finsler space-time in a way first discussed by Pirani \cite{Pirani1964,Rutz1998}, to wit
\begin{eqnarray}
\label{field equation}
0&=&\nonumber Ric\equiv R^\mu_{~\mu}\\
&=&\frac{1}{F^2}\left(2\frac{\partial G^\mu}{\partial x^\mu}-y^\alpha\frac{\partial^2 G^\mu}{\partial x^\alpha\partial y^\mu}+2G^\alpha\frac{\partial^2 G^\mu}{\partial y^\alpha\partial y^\mu}-\frac{\partial G^\mu}{\partial y^\alpha}\frac{\partial G^\alpha}{\partial y^\mu}\right)\ ,
\end{eqnarray}
where
\begin{equation}
\label{geodesic spray}
G^\mu=\frac{1}{4}g^{\mu\nu}\left(\frac{\partial^2 F^2}{\partial x^\alpha \partial y^\nu}y^\alpha-\frac{\partial F^2}{\partial x^\nu}\right)
\end{equation}
are called the geodesic spray coefficients \cite{Book by Bao}. $Ric$ is the Ricci scalar, which is given by the trace of the Ricci curvature tensor $R^{\lambda}_{~\mu}=\ell^{\kappa}R^{~\lambda}_{\kappa~\mu\nu}\ell^{\nu}$, where $\ell^{\mu}\equiv y^{\mu}/F$.
In terms of the torsion-free and almost metric-compatible Chern connection \cite{Chern1948,Chern1989}
\begin{equation}\label{Chern connection}
\Gamma^{\alpha}_{~\mu\nu}=\gamma^{\alpha}_{~\mu\nu}-g^{\alpha\lambda}\left(A_{\lambda\mu\beta}\frac{N^\beta_{~\nu}}{F}-A_{\mu\nu\beta} \frac{N^\beta_{~\lambda}}{F}+A_{\nu\lambda\beta}\frac{N^\beta_{~\mu}}{F}\right),
\end{equation}
the curvature of Finsler space is given as
\begin{equation}\label{Finsler curvature}
R^{~\lambda}_{\kappa~\mu\nu}=\frac{\delta
\Gamma^\lambda_{~\kappa\nu}}{\delta x^\mu}-\frac{\delta
\Gamma^\lambda_{~\kappa\mu}}{\delta
x^\nu}+\Gamma^\lambda_{~\alpha\mu}\Gamma^\alpha_{~\kappa\nu}-\Gamma^\lambda_{~\alpha\nu}\Gamma^\alpha_{~\kappa\mu},
\end{equation}
where $\frac{\delta}{\delta x^\mu}\equiv\frac{\partial}{\partial x^\mu}-N^\nu_{~\mu}\frac{\partial}{\partial y^\nu}$ and $\gamma^{\alpha}_{~\mu\nu}$ is the formal Christoffel symbols of the second kind with the same form of Riemannian connection. $N^\mu_{~\nu}$ in Eq.(\ref{Chern connection}) and Eq.(\ref{Finsler curvature}) are defined as $N^\mu_{~\nu}\equiv\gamma^\mu_{~\nu\alpha}y^\alpha-A^\mu_{~\nu\lambda}\gamma^\lambda_{~\alpha\beta}y^\alpha y^\beta$. $A_{\lambda\mu\nu}\equiv\frac{F}{4}\frac{\partial}{\partial y^\lambda}\frac{\partial}{\partial y^\mu}\frac{\partial}{\partial y^\nu}F^2$ is the Cartan tensor (regarded as a measurement of deviation from the Riemannian Manifold). The geodesic equation in Finsler space-time is given as \cite{Book by Bao}
\begin{equation}
\label{geodesic}
\frac{d^2x^\mu}{d\tau^2}+2G^\mu=0\, .
\end{equation}

It should be noted that in the Finslerian MOND, Eq.(\ref{field equation}) is simply a stipulation \cite{Lietal2012}. But its physical significance is very clear. In the geodesic deviation equation \cite{Book by Bao}, vanishing of the Ricci scalar $Ric$ implies that the geodesic rays are parallel to each other. This means that it is vacuum outside the gravitational source. Pfeifer \& Wohlfarth  have set up gravitational dynamics for Finsler space-time in terms of an action integral on the unit tangent bundle \cite{Pfeifer2012}. The stipulation $Ric=0$ here is compatible with their results of gravitational field equation\footnote{The gravitational vacuum field equation given by \cite{Pfeifer2012} is $g^{F\,ab}\bar\partial_a\bar\partial_b \mathcal{R}-\frac{6}{F^2} \mathcal{R}+2g^{F\,ab}\big(\nabla_aS_b+S_aS_b+\bar\partial_a\nabla S_b\big)=0$. The $S_a$-terms can be written as $S_a=\ell^{d}P_{d~ba}^{~b}$, where $\ell^d\equiv y^d/F$ and $P_{d~ba}^{~b}$ are the coefficients of the cross basis $dx \wedge \delta y/F$ \cite{Book by Bao}. Considering that $\mathcal{R}=R^a_{~ab}y^b=-R^a_{~dab}y^d y^b=F^2(\ell^d R^{~a}_{d~ab}\ell^b)=F^2(g^{ab}R_{ab})=F^2 Ric$ and dropping the $S_a$-terms, one can see that $Ric=0$ is one of the solutions of the above equation.}.

In Finsler geometry, the counterpart of the flat space-time in Riemann geometry (i.e. the Minkowski space-time) is called ``locally Minkowski space-time". In a locally Minkowski space-time, $F$ depends not on a local coordinate system $\{x^\mu\}$ but just on the induced tangent space coordinates $\{y^\mu\}$, i.e. $F=F(y)$. Using Eq.(\ref{field equation}), one can see that it is a solution of Finslerian vacuum field equation. Now let us assume that the metric is close to a locally Minkowski one $\eta_{\mu\nu}(y)$, i.e.
\begin{equation}\label{expand metric}
g_{\mu\nu}=\eta_{\mu\nu}(y)+h_{\mu\nu}(x,y)~~~~~~~~|h_{\mu\nu}|\ll1\ ,
\end{equation}
considering that the gravitational field $h_{\mu\nu}$ is stationary (thus all time derivatives of $h_{\mu\nu}$ vanishes) and the particle is moving very slowly (i.e. $GM/r\ll 1$). The lowering and raising of indices are carried out by $\eta_{\mu\nu}$ and its matrix inverse $\eta^{\mu\nu}$. Note that both \(\eta(y)\) and \(h(x,y)\) contain the spacetime anisotropy.

From the Finsler gravitational vacuum equation $Ric=0$ and taking into account of the matter, we find that
\begin{equation}\label{static field eq1}
\eta^{ij}\frac{\partial^2 h_{\alpha\beta}}{\partial x^i \partial x^j} +\mathcal{O}\left(h_{\mu\nu} \right)=-\kappa\left(T_{\alpha\beta}-\frac{1}{2}\eta_{\alpha\beta}T^\lambda_{~\lambda}\right).
\end{equation}
$h_{00},h_{nn}$ are terms of the same order as $GM/r$ and $T_{\alpha\beta}$ is the energy-momentum tensor of matter. One prerequisite for Eq.(\ref{static field eq1}) is that Finsler gravity should reduce to Einstein's general relativity if the Finsler metric $g_{\mu\nu}$ reduces to a Riemannian one. Thus we find from Eq.(\ref{static field eq1}) that
\begin{eqnarray}\label{static field eq2}
\eta^{ij}\frac{\partial^2 h_{00}}{\partial x^i \partial x^j}&=&-8\pi_F G\rho\eta_{00},\\
\label{static field eq3}
\eta^{ij}\frac{\partial^2 h_{nn}}{\partial x^i \partial x^j}&=&8\pi_F G\rho\eta_{nn},
\end{eqnarray}
where $\rho= T_{00}/\eta_{00}$ is the energy density of the gravitational source and $G$ is the gravitational constant in Newtonian dynamics. $\pi_F$ in Eq.(\ref{static field eq2}) and Eq.(\ref{static field eq3}) is given as
\begin{equation}
\pi_F\equiv\frac{3}{4}\int_{R=1}\sqrt{g}dx^1\wedge dx^2\wedge dx^3,
\end{equation}
where $g\equiv {\rm det}(\eta_{ij})$ is the determinant of $\eta_{ij}$ and ``$\wedge$" denotes the ``wedge product"\footnote{The symmetry of locally Minkowski space-time is different from that of Minkowski spacetime. The space length determined by the symmetry of locally Minkowski space-time is also different from the Euclidean length. So does the unit circle and its related quantity-$\pi$. Here, we denote the Finslerian $\pi$ by $\pi_F$, where ``$\wedge$" is the ``wedge product". For more details please refer to the book of \cite{Book by Chern}.}.
The solutions of Eq.(\ref{static field eq2}) and Eq.(\ref{static field eq3}) are given as
\begin{equation}\label{static field eq4}
h_{00}=-\frac{2GM}{R}\eta_{00},~~ h_{nn}=\frac{2GM}{R}\eta_{nn},
\end{equation}
where $R^2\equiv\eta_{ij}x^i x^j$. We find that the spacetime anisotropy is approximately determined by the locally Minkowski metric \(\eta\).

In the Newtonian limit, the geodesic equation Eq.(\ref{geodesic}) reduces to
\begin{eqnarray}\label{geodesic compo1}
\frac{d^2x^0}{d\tau^2}-\frac{\eta^{0i}}{2}\frac{\partial h_{00}}{\partial x^i}\frac{dx^0}{d\tau}\frac{dx^0}{d\tau}=0,\\
\label{geodesic compo2}
\frac{d^2x^i}{d\tau^2}-\frac{\eta^{ij}}{2}\frac{\partial h_{00}}{\partial x^j}\frac{dx^0}{d\tau}\frac{dx^0}{d\tau}=0.
\end{eqnarray}
Eq.(\ref{geodesic compo1}) implies that $\frac{dx^0}{d\tau}$ is a function of $h_{00}$. Since $|h_{00}|\ll1$, $\frac{dx^0}{d\tau}$ can be taken as a constant in Eq.(\ref{geodesic compo2}). Then we find from Eq.(\ref{geodesic compo2}) that
\begin{equation}\label{law of gravity}
\frac{d^2x^i}{{dx^0}^2}=-\frac{GM}{R^2}\frac{x^i}{R},
\end{equation}
where ${dx^0}^2=\eta_{00}dx^0 dx^0$. Thus, the dynamics of a test particle circling around a static gravitational point source with mass $M$ reads
\begin{equation}\label{eq:finsler-gravity}
  \frac{GM}{R^2}=\frac{v^2}{R}\ ,
\end{equation}
where $v$ is the circular rotation velocity of the test particle. Eq.(\ref{eq:finsler-gravity}) implies that the law of gravity in Finsler space-time is similar to that in Newtonian dynamics. The only difference is that the spatial distance in Minkowski space $r\equiv\sqrt{\delta_{ij}x_ix_j}$ is now replaced by that in Finsler space $R\equiv\sqrt{\eta_{ij}(y)x_ix_j}$, where $\delta_{ij}={\rm diag}(1,1,1)$ is the Euclidean metric and $\eta_{ij}(y)$ is the metric in locally Minkowski space. This is what we expect from Finslerian gravity, because the length difference is one of the most distinguishing features of Finsler geometry as compared to the Riemannian geometry. The spatial distances in the two different spacetimes can be related by \cite{Lietal2012}
\begin{equation}\label{eq:R-r-relarion}
  R=rf\big{(}v(r)\big{)},
\end{equation}
where $f\big{(}v(r)\big{)}$ is a function of $v$ and $r$, which depends on the Finsler structure $F$, or fundamentally speaking, the symmetry of the Finsler space-time. Here, the function \(f(v(r))\) characterizes all the possible anisotropy of the spacetime. It could be constrained by the galaxy rotation curves in section \ref{sec:mass-profile}. When \(f(v(r))=1\), the spacetime anisotropy vanishes and the Finsler spacetime reduces back to the Riemann one.

\section{Relations with the Milgrom's MOND}\label{sec:MMOND}

The Finslerian gravity Eq.(\ref{eq:finsler-gravity}) includes the Milgrom's MOND as its special case. If we take $f\big{(}v(r)\big{)}=\sqrt{1-(GMa_0/v^4)^2}$, Eq.(\ref{eq:finsler-gravity}) becomes
\begin{equation}\label{eq:finsler1}
  \frac{GM}{r^2}=\frac{v^2}{r}\mu\left(\frac{a}{a_0}\right),
\end{equation}
where $a\equiv v^2/r$ is the acceleration, $\mu(x)\equiv x/\sqrt{x^2+1}$ is the interpolation function and $a_0$ is a critical acceleration.
This is simply the Milgrom's MOND \cite{Milgrom1983a,Milgrom1983b}. It is noteworthy that \(f(v)\) is a function of the 3-velocity \(v\). This reveals that the time-space rotations are not preserved in the Finslerian MOND.

On the other hand, if we choose $f\big{(}v(r)\big{)}=[1+\sqrt{a_0r^2/GM}\,]^{-1}$, Eq.(\ref{eq:finsler-gravity}) becomes
\begin{equation}\label{eq:finsler2}
  v^2=\frac{GM}{r}+\sqrt{GMa_0}\ .
\end{equation}
The value of \(a_{0}\) determines the scale for the occurrence for the spacetime anisotropy. Only in the case \(GM/r^2<a_0\), the anisotropy of spacetime becomes significant. When the parameter \(a_0\) vanishes, the Finsler MOND reduces back to the Newtonian dynamics. Thus, \(a_0\) completely determines the spacetime anisotropy.

\begin{figure}
  \centering
 \includegraphics[width=12 cm]{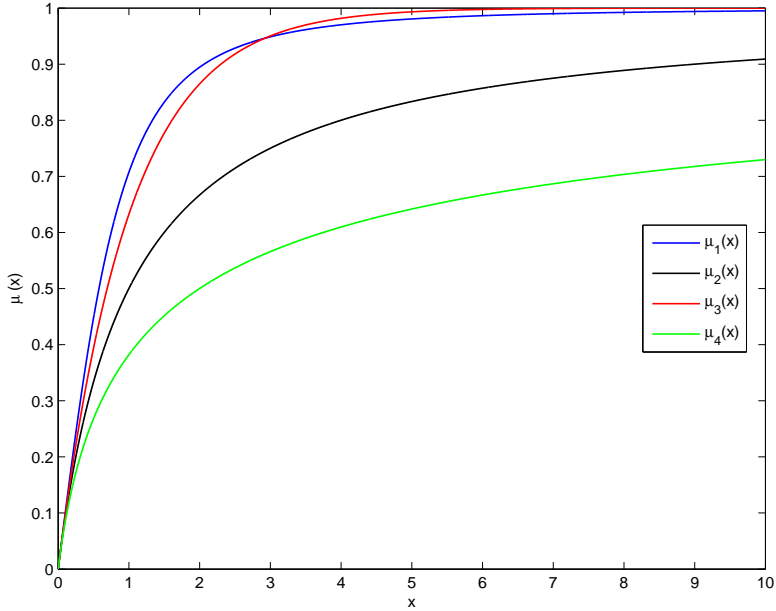}
 \caption{\small{Different interpolation functions. $\mu_1(x)=x/\sqrt{1+x^2}$ is the most commonly used interpolation function in the Milgrom's MOND. $\mu_2(x)=x/(1+x)$ and $\mu_3(x)=1-e^{-x}$. $\mu_4(x)=1+\frac{1}{2x}-\sqrt{\frac{1}{x}+\frac{1}{4x^2}}$ is the interpolation function in the Finslerian MOND.}}
  \label{fig:interp-function}
\end{figure}

Eq.(\ref{eq:finsler2}) can be rewritten as
\begin{equation}\label{eq:FMOND-velocity}
  v^2(r)=v_N^2\left(1+\frac{\sqrt{a_0r}}{v_N}\right),
\end{equation}
where $v_N\equiv\sqrt{GM/r}$ is the rotational velocity derived from Newtonian dynamics. This velocity has the same asymptotic behavior as that of the Milgrom's MOND
\begin{equation}\label{eq:limit-v}
  v(r)=\begin{cases}v_N(r)\quad \ \ & r\longrightarrow 0,\\
  (GMa_0)^{1/4}\quad \ \ & r\longrightarrow \infty.
  \end{cases}
\end{equation}
In fact, if we choose the interpolation function as\footnote{The interpolation function Eq.(\ref{eq:interp}) is singular at $x=0$, but it is easy to show that $\displaystyle\lim_{x\rightarrow 0}[\mu(x)/x]=1$.}
\begin{equation}\label{eq:interp}
  \mu(x)=1+\frac{1}{2x}-\sqrt{\frac{1}{x}+\frac{1}{4x^2}}\,,
\end{equation}
Eq.(\ref{eq:finsler1}) just becomes to Eq.(\ref{eq:finsler2}). For this reason, it is reasonable to call Eq.(\ref{eq:finsler2}) MOND, and We call it the Finslerian MOND in order to distinguish it from the Milgrom's MOND. We plot the different interpolation functions in Figure \ref{fig:interp-function}. We can see that the interpolation function used in the Finslerian MOND model increases much slower than others.

\section{The mass profile and the best-fit procedure}\label{sec:mass-profile}
To derive the theoretical rotational velocity in Newtonian theory, we should know the mass distribution of a galaxy. Generally, the luminosity mass in a galaxy mainly contains two components: gas (mainly HI and He) and stellar disk. Some early type galaxies may also contain a bulge in the center. However, the mass of the bulge is often much smaller than that of the stellar disk, and we will neglect it for simplicity. Because the dark matter has not yet be detected directly, we will not consider it here. Since the THINGS (The HI Nearby Galaxy Survey)\footnote{http://www.mpia-hd.mpg.de/THINGS/Data.html \cite{Walter2008}.} brought an unprecedented level of precision to the measurement of the rotation curves of galaxies, we choose the THINGS galaxies as the samples. The profile of neutral hydrogen (HI) is read directly from the THINGS data cube (robust weighted moment-0) using the task \textsc{ellint} of Groningen Image Processing System (\textsc{gipsy})\footnote{http://www.astro.rug.nl/$\sim$gipsy/.}. Assuming that HI locates in an infinitely thin disk, we calculate the rotational velocity contributed by HI ($v_{\rm HI}$) directly using the task \textsc{rotmod} of \textsc{gipsy}. More details of the THINGS galaxies can be found in reference \cite{Walter2008}.

The profile of stellar disk is derived from the photometric data. The surface brightness of the stellar disk can be fitted well by an exponential law \cite{Vaucouleurs1959,Freeman1970}
\begin{equation}\label{eq:brightness}
  I(r)=I(0)e^{-\frac{r}{h}},
\end{equation}
where $h$ is the scale length and $I(0)$ is the surface brightness at the center of galaxy. The integration of Eq.(\ref{eq:brightness}) gives the total luminosity
\begin{equation}
  L=2\pi I(0)h^2.
\end{equation}
If we assume that the mass-to-light ratio ($\Upsilon_*$) is a constant for each galaxy and the stellar disk is infinitely thin, the mass profile of the stellar disk can be written as
\begin{equation}\label{eq:surface-density}
  \Sigma(r)=\Sigma(0)e^{-\frac{r}{h}},
\end{equation}
where $\Sigma(0)=\Upsilon_{*}I(0)$ is the central surface mass density. The total mass of the stellar disk is given by
\begin{equation}\label{eq:disk-mass}
  M=\int_0^{\infty} 2\pi r \Sigma(r)dr=2\pi\Sigma(0)h^2.
\end{equation}
The mass-to-light ratio is given as
\begin{equation}\label{eq:mass-to-light}
  \Upsilon_{*}=\frac{M}{L}=\frac{\Sigma(0)}{I(0)}.
\end{equation}
The luminosity of a galaxy is often represented by its absolute magnitude.
The total luminosity $L$ is related to the absolute magnitude $\mathcal{M}$ as
\begin{equation}\label{eq:M-L-relation}
  \mathcal{M}-\mathcal{M}_{\odot}=-2.5{\rm log}_{10}\frac{L}{L_{\odot}},
\end{equation}
where $\mathcal{M}_{\odot}$ and $L_{\odot}$ are the absolute magnitude and the total luminosity of the sun, respectively.

The rotational velocity contributed by the exponential disk is derived by solving the Poisson equation with mass distribution Eq.(\ref{eq:surface-density}). The result is \cite{Freeman1970}
\begin{equation}\label{eq:disk-velocity}
  v_{*}(r)=\sqrt{\frac{M}{r}\gamma(r)},
\end{equation}
where
\begin{equation}
  \gamma(r)\equiv\frac{r^3}{2h^3}\left[I_0\left(\frac{r}{2h}\right)K_0\left(\frac{r}{2h}\right)- I_1\left(\frac{r}{2h}\right)K_1\left(\frac{r}{2h}\right)\right].
\end{equation}
Here, $I_{n}$ and $K_{n}$ ($n=0, 1$) are the $n$th order modified Bessel functions of the first and second kind, respectively. The Newtonian velocity due to the combined contributions of gas and stellar disk is given by
\begin{equation}\label{eq:newton-velocity}
  v_N=\sqrt{\frac{4}{3}v_{\rm HI}^2+v_{*}^2},
\end{equation}
where the factor 4/3 comes from the contribution of both helium (He) and neutral hydrogen (HI). Here we assume that the mass ratio of He and HI is $M_{\rm He}/M_{\rm HI}=1/3$. Any other gases are negligible compared to HI and He.

\begin{figure}
  \centering
 \includegraphics[width=14cm, height=20 cm]{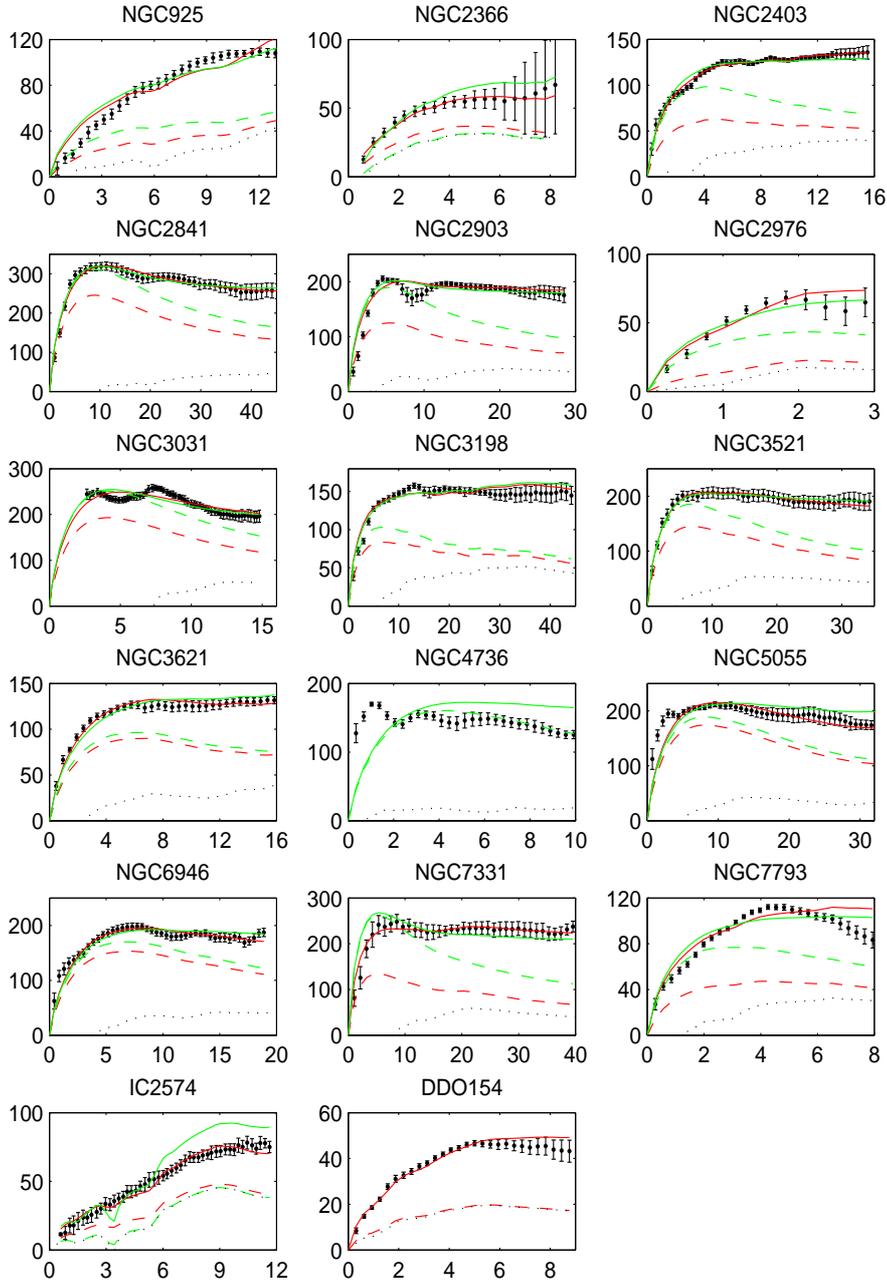}
  \caption{\small{The Finslerian MOND vs. the Milgrom's MOND. The $x$-axis is the distance in kpc, and the $y$-axis is the rotation velocity in km~s$^{-1}$. The black dotted curves are the contribution of gas (HI and He). The red solid and green solid curves are the results of the Finslerian MOND and the Milgrom's MOND, respectively.  The red dashed and green dashed curves are the Newtonian velocities calculated with the mass to light ratio deriving from the Finslerian MOND and the Milgrom's MOND, respectively. There is no solution found for NGC4736 in the Finslerian MOND fit, so as DDO154 in the Milgrom's MOND fit.}}
  \label{fig:bestfit}
\end{figure}

In the following discussions, we use Eq.(\ref{eq:FMOND-velocity}) and Eq.(\ref{eq:newton-velocity}) to fit the observed rotation curves.
The only two parameters are $\Sigma_0$ (or equivalently $\Upsilon_{*}$) and \(a_0\). The sample galaxies used in our paper are the same as that in references\cite{Chemin:2011mf,Mastache:2012ep}. These 17 galaxies are the subset of the 34 THINGS galaxies \cite{Walter2008}. We choose these galaxies because they have smooth, symmetric and high quality rotation curves extending to large distance. For convenience, we list their properties in Table \ref{tab:samples}.
\begin{table}
\begin{center}
\caption{\small{Properties of the sample galaxies. Column (1): galaxy names. Columns (2) and (3): galaxy centers in J2000.0 from \cite{Walter2008}. Columns (4) and (5): Inclinations  and position angles from \cite{Walter2008}. Column (6): systematic velocities from \cite{Blok:2008}. Column (7): distances from \cite{Walter2008}. Column (8): the scale lengths of optical disk from \cite{Blok:2008,Mastache:2012ep}. Column (9): mass of HI from \cite{Walter2008}. Column (10): apparent B-band magnitudes from \cite{Walter2008}. Column (11): absolute B-band magnitudes from \cite{Walter2008}. Column (12): luminosity calculated from column (11) using Eq.(\ref{eq:M-L-relation}).}}
\begin{tabular}[t]{cccccccccccc}
\hline\hline
(1) & (2)& (3) & (4) & (5) & (6) & (7) & (8) & (9) & (10) & (11)  &  (12) \\
 Names   &   RA  &  DEC  &  INCL  &  PA  &  $V_{\rm sys}$ &  $D$   &  $h$ &  $M_{\rm HI}$  & $m_B$  &  $M_B$  &  $L$ \\
 &  [h m s]  &  [$^{\circ}~^{\prime}~^{\prime\prime}$] &  [$^{\circ}$]  &  [$^{\circ}$]  & [km/s] & [Mpc] &  [kpc]  &  $[10^8 M_{\odot}]$ & [mag]  & [mag]  & [$10^{10} L_{\odot}$] \\
\hline
NGC925  & 02 27 16.5 & +33 34 44 & 66 & 287 & 546.3 & 9.2 & 3.30 & 45.8 &9.77 & \---20.04 & 1.614\\
NGC2366 & 07 28 53.4 & +69 12 51 & 64 & 40  & 104.0 & 3.4 & 1.76 & 6.49 &10.51 & \---17.17 & 0.115\\
NGC2403 & 07 36 51.1 & +65 36 03 & 63 & 124 & 132.8 & 3.2 & 1.81 & 25.8 & 8.11 & \---19.43 & 0.920\\
NGC2841 & 09 22 02.6 & +50 58 35 & 74 & 153 & 633.7 & 14.1 & 4.22 & 85.8 & 9.54 & \---21.21 & 4.742\\
NGC2903 & 09 32 10.1 & +21 30 04 & 65 & 204 & 555.6 &  8.9 & 2.40  & 43.5  & 8.82 & \---20.93 & 3.664\\
NGC2976 & 09 47 15.3 & +67 55 00 & 65 & 335 & 1.1   & 3.6 & 0.91 & 1.36 & 9.98 & \---17.78 & 0.201\\
NGC3031 & 09 55 33.1 & +69 03 55 & 59 & 330 & \---39.8 & 3.6 & 1.93 & 36.4 &7.07 & \---20.73 & 3.048\\
NGC3198 & 10 19 55.0 & +45 32 59 & 72 & 215 & 660.7 &  13.8 & 3.06 & 101.7 & 9.95 & \---20.75 & 3.105\\
NGC3521 & 10 05 48.6 & \---00 02 09 & 73 & 340 & 803.5 &  10.7 & 3.09 & 80.2 & 9.21 & \---20.94 & 3.698\\
NGC3621 & 11 18 16.5 & \---32 48 51 & 65 & 345 & 728.5 & 6.6 & 2.61 & 70.7 &9.06 & \---20.05 & 1.629\\
NGC4736 & 12 50 53.0 & +41 07 13 & 41 & 296 & 306.7 & 4.7 & 1.99 & 4.00 &8.54 & \---19.80 & 1.294\\
NGC5055 & 13 15 49.2 & +42 01 45 & 59 & 102 & 496.8 &  10.1 & 3.68 & 91.0 & 8.90 & \---21.12 & 4.365\\
NGC6946 & 20 34 52.2 & +60 09 14 & 33 & 243 & 43.7  & 5.9 & 2.97 & 41.5 &8.24 & \---20.61 & 2.729\\
NGC7331 & 22 27 04.1 & +34 24 57 & 76 & 168 & 818.3 &  14.7 & 2.41 & 91.3 & 9.17 & \---21.67 & 7.244\\
NGC7793 & 23 57 49.7 & \---32 35 28 & 50 & 290 & 226.2 & 3.9 & 1.25 & 8.88 & 9.17 & \---18.79 & 0.511\\
IC2574  & 10 28 27.7 & +68 24 59 & 53 & 56  & 53.1  & 4.0 & 2.56 & 14.8 & 9.91 & \---18.11 & 0.273\\
DDO154  & 12 54 05.9 & +27 09 10 & 66 & 230 & 375.9 & 4.3 & 0.72 & 3.58 & 13.94 & \---14.23 & 0.008\\
\hline
\end{tabular}\label{tab:samples}
\end{center}
\end{table}
In order to constrain the parameters, we define the chi-square as
\begin{equation}\label{eq:chi-aquare}
  \chi^2=\sum_{i=1}^n \frac{[v_i^{\rm obs}-v^{\rm th}(r_i)]^2}{\sigma_i^2},
\end{equation}
where $v_i^{\rm obs}$ is the observed rotation velocity, $v^{\rm th}(r_i)$ is the theoretical velocity at radius $r_i$ calculated from Eq.(\ref{eq:FMOND-velocity}), and $\sigma_i$ is the uncertainty of $v_i^{\rm obs}$. Then we employ the least-square method to minimize Eq.(\ref{eq:chi-aquare}). The results are plotted in Figure \ref{fig:bestfit} with red curves, and the values of parameters are listed in Table \ref{tab:bestfit}. In order to make a comparison between the Finslerian MOND and the Milgrom's MOND, we also plot the fitting results of the latter in Figure \ref{fig:bestfit} with green curves, and list the fitting parameters in Table \ref{tab:bestfit} within brackets.

\begin{table}
\begin{center}
\caption{\small{ The best-fit parameters. The numbers with and without brackets are the best-fit parameters for the Milgrom's MOND and the Finslerian MOND, respectively. The free parameters are $\Sigma_0$ and $a_0$ for the Finslerian MOND fit. For the Milgrom's MOND fit, the only one free parameter is $\Sigma_0$, while $a_0$ is fixed at $1.2 \times 10^{-10} \rm{m~s}^{-2}$. $M_{\rm disk}$ is the mass of the stellar disk calculated from Eq.(\ref{eq:disk-mass}). $M/L$ is the mass-to-light ratio calculated from Eq.(\ref{eq:mass-to-light}). $\chi^2$/dof is the reduced chi-square.
There is no solution found for NGC4736 in the Finslerian MOND fit, so as DDO154 in the Milgrom's MOND fit. The numbers after ``$\pm$" are the $1 \sigma$ errors of the corresponding parameters.}}
\begin{tabular}[t]{c|ccccc}
\hline\hline
& $\Sigma_0$ & $a_0$ & $M_{\rm disk}$ & $M/L$ & $\chi^2$/dof\\
& $[M_{\odot}~\rm{pc}^{-2}]$ & $[10^{-10} \rm{m~s}^{-2}]$ & $[10^{10} M_{\odot}]$ & $[M_{\odot}/L_{\odot}]$ & \\ \hline
NGC925  & $22.19\pm 7.24$   & $1.564\pm 0.379$ & $0.15\pm 0.05$ & $0.09\pm 0.03$ & 5.64\\
        & ($52.79\pm 5.68)$ & $(1.2\pm 0.0)$ & ($0.36\pm 0.04$)  & ($0.22\pm 0.02$) & (5.90)\\ \hline
NGC2366 & $22.92\pm 5.96$   & $0.186\pm 0.050$ & $(4.46\pm 1.16)\times 10^{-2}$ & $0.39\pm 0.10$ & 0.25\\
        & ($1.29\pm 1.40)$ & $(1.2\pm 0.0)$ & (($2.51\pm 2.72)\times 10^{-3}$)  & ($0.02\pm 0.02$) & (1.21)\\ \hline
NGC2403 & $189.18\pm 7.38$   & $1.871\pm 0.081$ & $0.39\pm 0.02$ & $0.42\pm 0.02$ & 0.77\\
        & ($506.76\pm 9.12)$ & $(1.2\pm 0.0)$ & ($1.04\pm 0.02)$  & ($1.13\pm 0.02$) & (2.46)\\ \hline
NGC2841 & $1409.26\pm 68.89$ & $0.954\pm 0.128$ & $15.77\pm 0.77$ & $3.33\pm 0.16$ & 0.79\\
        & ($2256.69\pm 33.84$)  & $(1.2\pm 0.0)$  & ($25.25\pm 0.38$)  & ($5.32\pm 0.08$) & (0.79)\\ \hline
NGC2903 & $628.07\pm 69.65$ &  $1.968\pm 0.345$ & $2.27\pm 0.25$ & $0.62\pm 0.07$ & 4.16\\
        & ($1464.27\pm 49.30$)  & $(1.2\pm 0.0)$ & ($5.30\pm 0.18$)  & ($1.45\pm 0.05$) & (5.41)\\ \hline
NGC2976 & $22.97\pm 17.01$   & $6.278\pm 3.913$ & $(1.20\pm 0.89)\times 10^{-2}$ & $0.06\pm 0.04$ & 2.04\\
        & ($173.70\pm 25.21)$ & $(1.2\pm 0.0)$ & ($(9.04\pm 1.31)\times 10^{-2}$)  & ($0.45\pm 0.07$) & (2.49)\\ \hline
NGC3031 & $1907.94\pm 152.74$   & $1.159\pm 0.284$ & $4.47\pm 0.36$ & $1.47\pm 0.12$ & 2.68\\
        & ($3226.14\pm 62.57)$ & $(1.2\pm 0.0)$ & ($7.55\pm 0.15$)  & ($2.48\pm 0.05$) & (3.92)\\ \hline
NGC3198 & $222.67\pm 19.28$  & $0.963\pm 0.111$ & $1.31\pm 0.11$ & $0.42\pm 0.04$ & 3.18\\
        & ($342.57\pm 12.72$)  & $(1.2\pm 0.0)$  & ($2.02\pm 0.07$)  & ($0.65\pm 0.02$) & (2.85)\\ \hline
NGC3521 & $660.82\pm 22.62$ & $0.943\pm 0.064$ & $3.96\pm 0.14$ & $1.07\pm 0.04$ & 0.21\\
        & ($1090.69\pm 13.04$)  & $(1.2\pm 0.0)$  & ($6.54\pm 0.08$)  & ($1.77\pm 0.02$) & (0.20)\\ \hline
NGC3621 & $282.70\pm 18.44$   & $0.507\pm 0.075$ & $1.21\pm 0.08$ & $0.74\pm 0.05$ & 0.91\\
        & ($330.09\pm 12.08)$ & $(1.2\pm 0.0)$ & ($1.41\pm 0.05$)  & ($0.87\pm 0.03$) & (2.49)\\ \hline
NGC4736 &   $\times$  & $\times$ & $\times$ & $\times$ & $\times$\\
        & ($1259.82\pm 176.55)$ & $(1.2\pm 0.0)$ & ($3.13\pm 0.44$)  & ($2.42\pm 0.34$) & (44.83)\\ \hline
NGC5055 & $801.72\pm 57.44$ & $0.298\pm 0.088$ & $6.82\pm 0.49$ & $1.56\pm 0.11$ & 2.77\\
        & ($945.69\pm 31.04$) & $(1.2\pm 0.0)$ &($8.05\pm 0.26$)  & ($1.84\pm 0.06$) & (4.77)\\ \hline
NGC6946 & $758.23\pm 42.80$   & $0.393\pm 0.075$ & $4.20\pm 0.24$ & $1.54\pm 0.09$ & 0.97\\
        & ($942.95\pm 20.54)$ & $(1.2\pm 0.0)$ & ($5.23\pm 0.11$)  & ($1.92\pm 0.04$) & (1.69)\\ \hline
NGC7331 & $735.94\pm 65.27$ & $3.742\pm 0.355$ & $2.69\pm 0.24$ & $0.37\pm 0.03$ & 0.37\\
        & ($2829.24\pm 144.04$)  & $(1.2\pm 0.0)$  & ($10.32\pm 0.53$) & ($1.43\pm 0.07$) & (2.04)\\ \hline
NGC7793 & $134.02\pm 32.49$   & $2.638\pm 0.738$ & $0.13\pm 0.03$ & $0.26\pm 0.06$ & 7.52\\
        & ($450.04\pm 26.93)$ & $(1.2\pm 0.0)$ & ($0.44\pm 0.03$)  & ($0.86\pm 0.05$) & (10.11)\\ \hline
IC2574  & $12.20\pm 2.25$   & $0.178\pm 0.021$ & $(50.24\pm 9.26)\times 10^{-3}$ & $0.18\pm 0.03$ & 1.10\\
        & ($0.63\pm 1.18)$ & $(1.2\pm 0.0)$ & (($2.59\pm 4.86)\times 10^{-3}$)  & ($0.01\pm 0.02$) & (9.70)\\ \hline
DDO154  & $3.29\pm 1.04$ & $0.558\pm 0.028$ & $(1.07\pm 0.34)\times 10^{-3}$ & $0.13\pm 0.04$ & 1.33\\
        & ($\times$) & $(\times)$  &  ($\times$)  & ($\times$) & ($\times$)\\ \hline
\end{tabular}\label{tab:bestfit}
\end{center}
\end{table}

From Table \ref{tab:bestfit} and Figure \ref{fig:bestfit}, we can see that the Finslerian MOND fits the data as well as the Milgrom's MOND.
For some galaxies, such as NGC2403, NGC2841, NGC2903, NGC3521, the theoretical curves of the two MOND theories are so close to each other that they seem to overlap. The mass-to-light ratio obtained from the Finslerian MOND is often smaller than that from Milgrom's MOND. For NGC4736, no solution was found in the Finslerian MOND fit, and the Milgrom's MOND fit is extremely poor. For the gas-rich dwarf galaxy DDO154, no solution was found in the Milgrom's MOND fit. In the Finslerian MOND fit, the critical accelerations $a_0$ for different galaxies shows a fluctuation, but they are all in the order of magnitude $10^{-10}~{\rm m}~{\rm s}^{-2}$. The value of $a_0$ for NGC2976 is much higher than others. This may because of the poor quality of the data\----the observed velocity extends to only about 3 kpc and shows a trend of decline at the outmost point. Another interesting feature is that the value of $a_0$ for the dwarf galaxies (such as NGC2366,  IC2574 and DDO154) is smaller than that of others. The average value of $a_0$ (except for NGC2976 and DDO4736) is \(\bar{a}_0\approx cH_0/2\pi \sim 1.2\times 10^{-10}~{\rm m}~{\rm s}^{-2}\), which is in good agreement with that of the Milgrom's MOND. Here, \(c=3\times10^8\) m s\(^{-1}\) is the speed of light and \(H_0=75\) \({\rm km}~{\rm s}^{-1}~{\rm Mpc}^{-1}\) is the Hubble constant. Most importantly, the magnitude of \(a_0\) gives the scale for the occurrence of the spacetime anisotropy. When the Newtonian acceleration \(GM/r^2\) is smaller than \(a_0\), the anisotropic effect of the spacetime becomes significant.

\section{Discussions and conclusions}\label{sec:discussion}
The spacetime anisotropy has been tested by physical observations at different scales. It is related with certain privileged axes in the spacetime. As was mentioned in the introduction, several evidences were found for the preferred axes at the large scales in the universe,
even though their confidence levels are not very significant. On the earth, the spacetime anisotropy behaves as the Lorentz invariance violation (LIV) and the CPT violation, although the levels of Lorentz and CPT violations have been constrained severely, see review in reference\cite{Data tables for Lorentz and CPT violation}. At the galactic scales, however, the spacetime anisotropy could account for the galaxy rotation curves as was addressed in this paper. Or equivalently, the galaxy rotation curves give a possible signal to the spacetime anisotropy. Maybe there exist certain different physical effects at the galactic scales from those at other scales. It is noteworthy that the spacetime anisotropy could be accounted by Finsler geometry. The reason is that the Finsler spacetime is intrinsically anisotropic. For the D-dimensional spacetime, there are only \(D(D-1)/2+1\) Killing vectors in the Finsler spacetime. Thus, Finsler geometry could be a candidate to deal with issues of the privileged axes in the spacetime.

It should be mentioned that Grumiller has derived an effective gravitational potential including a Newton term ($\propto r^{-1}$) plus a Rindler term ($\propto r$) \cite{Grumiller:2010bz}. According to Grumiller's modified gravity, the rotation velocity of a test particle moving in the gravitational potential of a point particle with mass $M$ takes the form $v^2(r)=GM/r+ar$ \cite{Grumiller:2010bz}, where $a$ is the Rindler acceleration in order of magnitude $10^{-10}$ m s$^{-2}$. The Rindler acceleration may be a constant, or a function of $r$. In the case of $a$ being a constant, it was showed that Grumiller's theory can well reproduce the observations in the range $0\lesssim r \lesssim 40$ kpc, with the Rindler acceleration converges to an average value $\bar{a}\approx0.3\times 10^{-10}$ m s$^{-2}$ \cite{Lin:2012zh}. However, the theoretical curves show a tendency of sharply arising beyond the data range. Mastache et al. pointed out that Grumiller's model will fail if more galaxies are considered: the Rindler acceleration shows a high spread for different galaxies and in many cases $\chi^2/{\rm dof}\gg 1$ \cite{Mastache:2012ep}. As was mentioned in reference \cite{Grumiller:2010bz}, $a$ is not necessary to be a constant, it may be a function of $r$. If we assume that $a$ is inversely proportional to $r$ and choose the proportional factor to be $\sqrt{GMa_0}$, ie., $a=\sqrt{GMa_0}/r$, Grumiller's velocity reduces to that of Finslerian MOND.

In a previous paper \cite{Lietal2012}, we have already showed that the Finslerian MOND can well explain the mass discrepancy problem of Bullet Cluster 1E0657-558, which is a challenge to Milgrom's MOND \cite{Angus2006,Angus2007}. If we take the dipole and quadrupole effects into consideration, the function \(f(v(r),\theta)\) in Eq.(\ref{eq:R-r-relarion}) takes the form
\begin{equation}\label{eq:qd}
  f^{-1}(v(r),\theta)=1+\sqrt{\frac{a_0r^2}{GM}}\left(1+\frac{\sqrt{GMa_0}}{a^2}\cos\theta\exp(-r/c)+\frac{GMa_0}{b^4}\cos^2\theta\exp(-r/c)\right),
\end{equation}
where $a$ and $b$, which have the dimension of velocity, characterize the strongness of the dipole and quadrupole effects, respectively. The exponential term is needed in order to keep the Tully\---Fisher relation. The first term on the right-hand-side of Eq.(\ref{eq:qd}) is the classical Newtonian term, the second term is the Finslerian correction, the third and fourth terms are the contributions of dipole and quadrupole effects, respectively. In the leading order approximation, the convergence $\kappa$-map becomes to $\kappa_F=\kappa_G f^{-1}(v(r),\theta)$, where $\kappa_G$ is the convergence $\kappa$-map derived from the classical general relativity \cite{Lietal2012}. This model can reproduce the general feature of the Bullet Cluster 1E0657-558: (1) the observed $\kappa$-map is much larger than that deduced from general relativity, (2) the gravitational center is a few kpc away from the mass center, (3) the peak and the sub-peak of the $\kappa$-map are asymmetric, (4) the theoretical temperature is well consistent with the observation.

Another challenge to the Milgrom's MOND involves the gravitational lensing \cite{Ferreras2008,Ferreras2009,Ferreras2012}. Ferreras et al. studied six strong gravitational lensing early-type galaxies from the CASTLES sample and showed that a significant amount of dark matter is need even in the framework of Milgrom's MOND \cite{Ferreras2008}. They also argued that if the interpolation function $\mu(x)$ increases with $x$ slowly enough, the need of dark matter may diminish. The interpolation function in Eq.(\ref{eq:interp}) happens to satisfy this condition. Thus we may expect that the Finslerian MOND can explain the gravitational lensing without dark matter. The details of this problem are left to another publication.

In this paper, we derived a modified formula for the galaxy rotational velocity, which has the same asymptotic behavior as the Milgrom's MOND, in the framework of Finslerian geometry. The best-fit to the 17 THINGS galaxies showed that the Finslerian MOND could well reproduce the observed data. We found that the critical accelerations $a_0$ of 15 out of the 17 galaxies (except for NGC2976 which has a large $a_0$ due to the poor data, and NGC4736 which has no solution) have an average value $\bar{a}_0\approx 1.2\times 10^{-10}$ m s$^{-2}$, which is in accordance with that of the Milgrom's MOND. In the Finslerian MOND, the parameter \(a_0\) determines the critical scale of the occurrence of the spacetime anisotropy. Thus, the anisotropic effect of the spacetime becomes significant when the Newtonian acceleration \(GM/r^2\) is smaller than the critical acceleration \(a_0\).

\begin{acknowledgements}
We are grateful to Y. G. Jiang for useful discussions. This work has been funded in part by the National Natural Science Fund of China under Grant No. 11075166 and No. 11147176.
\end{acknowledgements}



\end{document}